**Universal Density Control of Surface Reconstruction in Two-Dimensional Metals**

Jinyuan Yang[1] and Xiaoliang Zhong[1,*]

We identify a unified thickness criterion for surface reconstruction in ultrathin metal sheets. The critical thickness is governed by the change in surface atomic density upon reconstruction: thinning always favors structural changes that increase this density. Thus, thinning suppresses the (1 × 2) reconstruction of 5d noble – metal (110) sheets but promotes the quasi-hexagonal reconstruction of 4d noble–metal (001) sheets. Density-functional calculations validate these trends and show only minor corrections from surface stress and quantum-oscillation effects.

[1]School of Energy and Power Engineering, Huazhong University of Science and Technology, 1037 Luoyu Road, Wuhan, China

*xzhong@hust.edu.cn

Metal surface reconstruction can profoundly affect the stability and functionality of metal surfaces [1-3]. Due to their limited size, metal nanostructures may exhibit surface reconstruction behaviors different from those of bulk surfaces. Among them, metal nanoparticles have been the most extensively studied [4-8]. For example, whereas the initially unreconstructed bulk Au(111) surface does not reconstruct unless a relatively high temperature is applied, the (111) facets of Au nanoparticles readily reconstruct at 300 K due to the presence of edges and vertices [8]. Recently, noble-metal nanosheets have emerged as a promising class of materials in catalysis, bioimaging, and sensing [9-12]. However, while it is known that many noble metal bulk surfaces do reconstruct [1], whether 2D noble metals exhibit similar reconstruction behavior remains largely unexplored, which is largely because high-quality 2D noble-metal sheets have only recently become experimentally accessible. Huang et al. reported that when the thickness of Au sheets was reduced to 2.4 nm, the hcp phase was obtained rather than the conventional fcc phase [13]. Nevertheless, most nanosheets of fcc noble metals maintain the fcc phase [10,12], and whether their surfaces reconstruct remains unknown. Zhao et al. discovered surface reconstruction of Pd nanosheets induced by adsorbed CO [14]. In this work, we focus on the surface reconstruction of clean noble-metal nanosheets, and the insights gained are expected to deepen the understanding of adsorbate-induced surface reconstructions.

Clearly, the lack of knowledge about surface reconstruction of 2D noble metals significantly hinders the understanding and prediction of their properties. In the present work, we study the energetics of fcc noble-metal nanosheets with either the (110) or (001) surface exposed, aiming to identify the key factors governing the surface reconstruction of 2D metals. For bulk surfaces, which have an infinite number of atomic layers underneath, surface energy should be used to assess their thermodynamic stability. For nanosheets, which contain only a finite number of atomic layers, surface energy is no longer well defined. Moreover, the unreconstructed and reconstructed nanosheets may have different surface atomic densities. In this work, for each nanosheet of a given thickness, we calculate and compare the energies per atom of the unreconstructed and reconstructed structures to assess their relative stability, which is a commonly used method for evaluating the stabilities of elemental nanomaterials with different numbers of atoms [15,16]. For bulk surfaces, it is known that both the (110) and (001) surfaces of the 5d noble metals (Ir, Pt, and Au) reconstruct to form (111)-like structures, thereby effectively reducing the surface energies, whereas those of the 4d noble metals (Rh, Pd, and Ag) do not reconstruct [1,3]. Notably, while the reconstructed (110)-(1×2) surfaces of the 5d noble metals halve the surface atomic density, their reconstructed (001) surfaces increase it [17-19]. We will show that the variation in surface atomic density plays a critical role in changing the reconstruction tendency of nanosheets.

For an unreconstructed slab consisting of $l$ atomic layers with $m$ atoms in each layer, the total energy ($E_{UN}$) is given by

$$E_{UN} = lmE_{bulk} + 2mE_{S-UN} \tag{1}$$

where $E_{bulk}$ is the energy per atom in the bulk and $E_{S-UN}$ is the surface energy of the unreconstructed surface per (1×1) area. The factor of 2 accounts for the two equivalent surfaces of the slab. Eq. (1) becomes asymptotically exact for thick slabs, whereas for nanosheets with thicknesses of a few nanometers or less, quantum-size oscillations and strain effects due to surface stress are expected to affect the energies. When surface reconstruction is included, we assume that each reconstructed surface layer contains $n$ atoms over the same lateral area; $n$ may be less than, equal to, or greater than $m$. Similarly, the total energy of the reconstructed system ($E_{RE}$) can be

written as

$$E_{RE} = ((l-2)m + 2n)E_{bulk} + 2mE_{S-RE} \tag{2}$$

where $E_{S-RE}$ is the surface energy of the reconstructed surface per (1×1) surface area. The *relative stability* is determined by the difference in the average energy per atom ($\Delta E$),

$$\Delta E = \frac{E_{RE}}{(l-2)m+2n} - \frac{E_{UN}}{lm} \tag{3}$$

A negative $\Delta E$ indicates that the reconstructed nanosheet is energetically favored, whereas a positive $\Delta E$ indicates that the unreconstructed one is favored. Using Eqs. (1) and (2), $\Delta E$ can be rewritten as

$$\Delta E = \frac{2}{l((l-2)m+2n)}(lm\Delta E_S + 2(m-n)E_{S-UN}) \tag{4}$$

where $\Delta E_S$ is the surface energy difference between the reconstructed and unreconstructed surfaces,

$$\Delta E_S = E_{S-RE} - E_{S-UN} \tag{5}$$

$\Delta E_S < 0$ if the reconstructed surface is more stable and $\Delta E_S > 0$ otherwise. Since both the top and bottom surface layers are reconstructed, at least one bulk-like layer is required ($l \geq 3$) and $(l-2)m + 2n$ is always positive. Setting $\Delta E = 0$ in Eq. (4) yields the critical thickness $l_c$ (in units of atomic layers) at which the reconstruction tendency changes,

$$l_c = 2(\frac{n}{m} - 1)\frac{E_{S-UN}}{\Delta E_S} \tag{6}$$

Note that $E_{S-UN}$ is always positive while $\Delta E_S$ can be either positive or negative. Thus, depending on the signs of $(\frac{n}{m} - 1)$ and $\Delta E_S$, $l_c$ may be negative, zero, or positive. When $l_c > 0$, $\Delta E$ crosses zero at $l = l_c$; that is, the reconstruction tendency changes at this thickness. When $l_c \leq 0$, the reconstruction tendency is not expected to change with the sheet thickness.

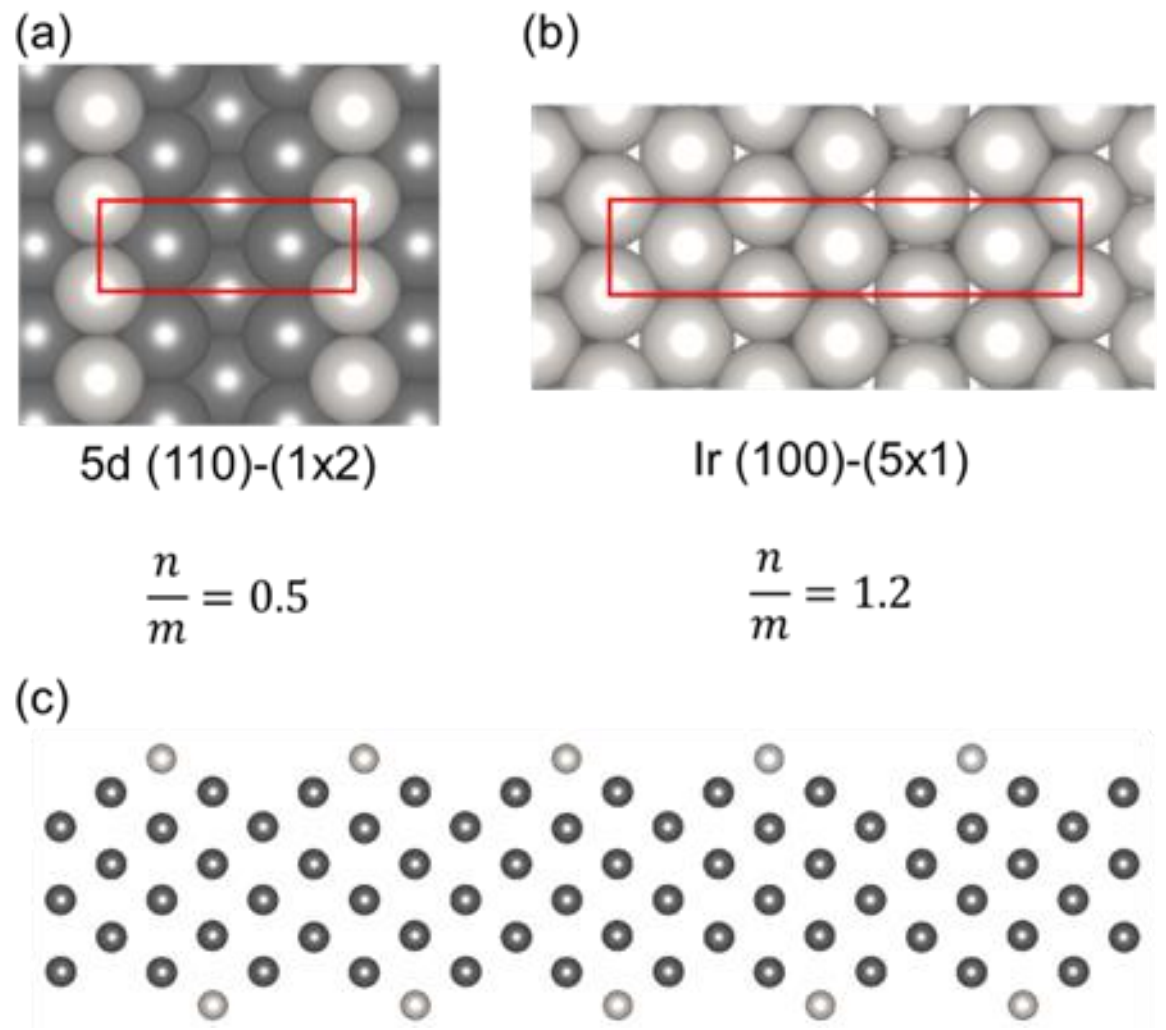

FIG. 1. (a) Atomic structure of the (110)-(1×2) reconstruction of fcc metals (top view). Atoms in the reconstructed surface layer are highlighted in light gray for clarity. *n* and *m* are the numbers of atoms in the reconstructed surface layer and in a bulk layer over the same area, respectively. (b) Atomic structures of the Ir(001)-(5×1) reconstruction (top view). The red rectangles in (a) and (b) mark the unit cells. (c) Atomic structure of an eight-layer (110) nanosheet with both surfaces exhibiting the (1×2) reconstruction (side view).

We now turn to prototypical noble metal surface reconstructions to highlight the critical role of surface atomic density variation in determining the reconstruction tendency of nanosheets. As shown in Fig. 1(a), the (110)-(1×2) missing-row reconstruction of 5d noble metals halves the surface atomic density relative to the bulk-terminated (110) surface, whereas 4d noble metal (110) surfaces do not reconstruct. In addition to the (1×2) reconstruction, other patterns have been reported for Ir(110) [20]. In this work, we focus on the (1×2) missing-row reconstruction for consistency with the cases of Pt and Au. To ascertain whether the unreconstructed or the reconstructed nanosheets are energetically more stable, we compute the average energy for two sets of structures via density functional theory (DFT) methods [21,22]. In one set, noble-metal nanosheets with both surfaces unreconstructed are generated and fully optimized; in the other set, nanosheets with both surfaces reconstructed (see Fig. 1(c)) are likewise fully optimized. DFT calculations are performed using the plane-wave technique implemented in the Vienna ab initio Simulation package (VASP) [23] with the PBEsol functional [24], which generally reproduces transition metal surface properties well [25]. The plane-wave cutoff energy is set to 400 eV in all computations. The convergence of energy is set as $10^{-6}$ eV. 16×16×1 and 16×8×1 Monkhorst–Pack k-point meshes are set for the unreconstructed and (1×2)-reconstructed (110) sheets, respectively, while 18×18×1 and 4×18×1 meshes are set for the unreconstructed and (5×1)-reconstructed (001) sheets, respectively.

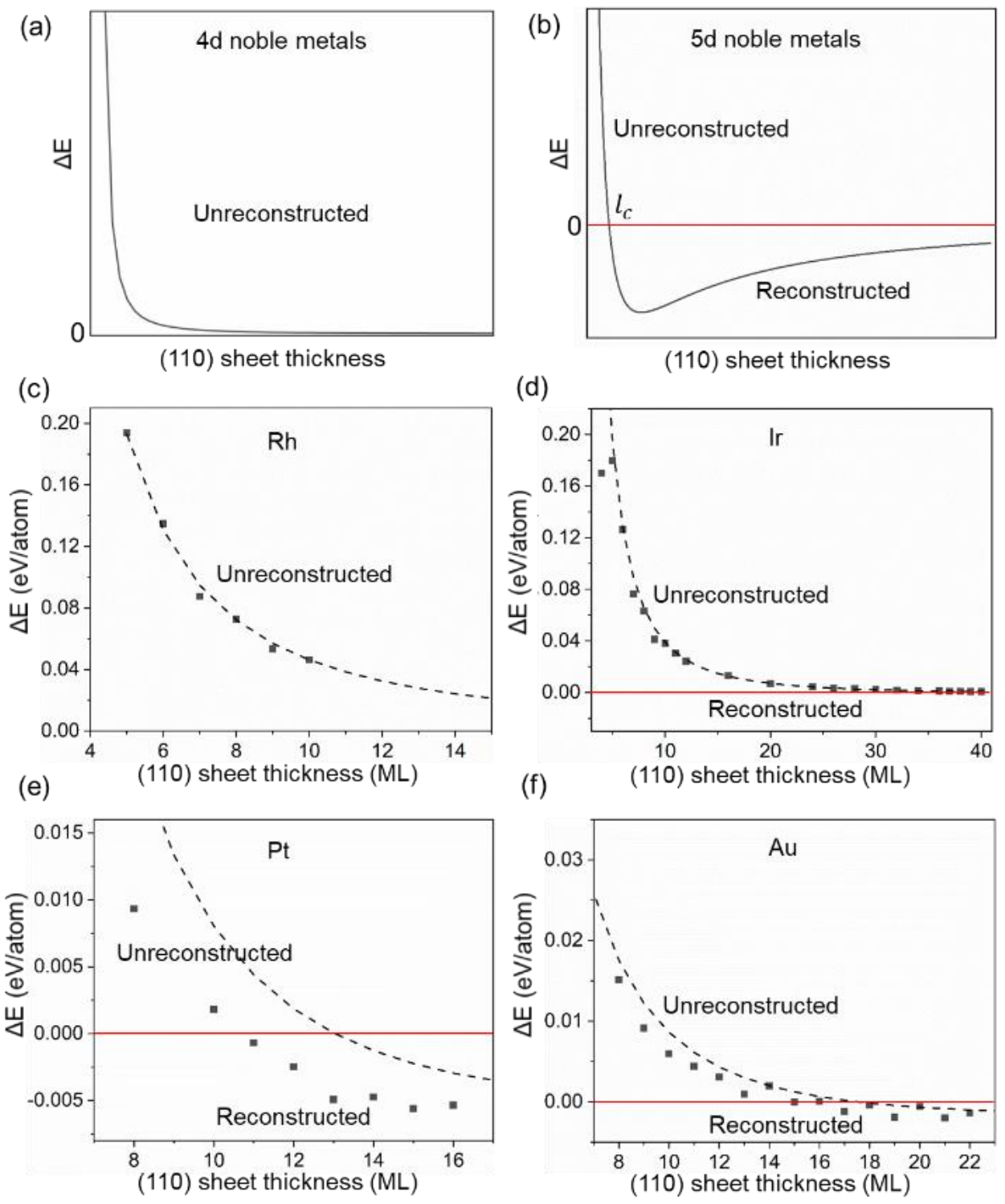

FIG. 2. (a), (b) Schematic energy difference ($\Delta E$) between (1 × 2)-reconstructed and unreconstructed structures as a function of (110) sheet thickness for 4d and 5d noble metals, respectively. $\Delta E > 0$ indicates that the unreconstructed structures are more stable, and vice versa. (c-f) $\Delta E$ versus sheet thickness for Rh, Ir, Pt and Au, respectively. The black dots are values obtained directly from total-energy calculations of fully optimized 2D structures, while the dashed curves are derived from surface energies using Eq. (4).

For 4d noble metals, bulk (110) surfaces do not reconstruct; that is, $\Delta E_S > 0$ (see Eq. (5)). Since $\frac{n}{m} = 0.5$, Eq. (6) yields $l_c < 0$. Therefore, as discussed above, the reconstruction tendency of 4d noble-metal nanosheets (i.e., remaining unreconstructed) is not expected to change with sheet thickness. As Fig. 2(a) shows, the average energy difference ($\Delta E$) approaches zero for thick sheets where surface effects become negligible. As sheet thickness decreases, $\Delta E$ increases because of the larger specific surface area. We plot the case of Rh in Fig. 2(c), which shows that $\Delta E$ values calculated from fully optimized structures (black dots) are very close to those derived from the surface energies (Eq. (4); dashed line). This indicates that quantum oscillations and lattice strain, which would otherwise cause $\Delta E$ to deviate from the derived values, have only minor effects. In the case of 5d metals, $\Delta E_S < 0$, so $l_c > 0$. Consequently, the reconstruction tendency is expected to change at $l = l_c$ (Fig. 2(b)). That is, for sufficiently thin ($l < l_c$) 5d noble-metal nanosheets, unreconstructed structures are expected to be more stable than the reconstructed ones. Compared with Ir (Fig. 2(d)), Pt and Au (110) sheets show a larger deviation between the directly calculated $\Delta E$ values and the derived ones. The lower values of the calculated $\Delta E$ originate from the greater surface stresses of the reconstructed surfaces [26], which, in turn, cause more significant lattice contraction and a larger energy decrease of the reconstructed nanosheets. Moreover, compared with Ir, Pt and Au have much lower Young's moduli. Therefore, lattice contraction and energy decrease are more significant for these two metals. We also observe that for all metals shown in Fig. 2 (Rh, Ir, Pt, and Au), nanosheets initially exposing (110) surfaces – whether unreconstructed, reconstructed, or both – transform into those exposing (001) surfaces when sufficiently thin. Similar phase transformations have been reported in fcc metal nanowires [27].

To further evaluate the effects of surface stress on nanosheet reconstruction, we compare in Table I the derived critical thickness ($l_c$ in Eq. (6)) with the actual critical thickness ($l_c^*$), the latter defined as the nanosheet thickness at which $\Delta E$ is closest to zero. For Ir (110), $l_c = 55.2$ layers, while our $\Delta E$ calculations are limited to sheets up to 40 layers thick, beyond which electronic convergence becomes rather slow. At this thickness, $\Delta E > 0$ (although being small, Fig. 2(d)); therefore, the actual critical thickness $l_c^* > 40$ layers. As Table I shows, for both Pt and Au $l_c^*$ is slightly smaller than $l_c$, which is attributed to the lower actual $\Delta E$ values compared with the derived ones, as discussed above (Figs. 2(e) and 2(f)). The close agreement between $l_c^*$ and $l_c$ indicates that the effects of surface stress on nanosheet reconstruction are small.

TABLE I. Comparison of the derived critical thickness $l_c$ (from surface energies) with the actual critical thickness $l_c^*$. The values of $l_c^*$ are obtained from total-energy calculations of fully optimized 2D nanosheets; therefore, effects such as surface stress and quantum oscillations are taken into account. The (1×2) reconstruction is considered for the (110) sheets, while the (5×1) reconstruction is considered for the (001) sheets.

| | Ir (110) | Pt (110) | Au (110) | Rh (001) | Pd (001) | Ag (001) |
|---|---|---|---|---|---|---|
| $l_c$ | 55.2 | 13.0 | 17.6 | 5.7 | 4.5 | 3.0 |
| $l_c^*$ | >40 | 11 | 15 | 6 | 5 | 4 |

We have shown that for 5d noble-metal (110) nanosheets, $\frac{n}{m} < 1$ and $\Delta E_S < 0$, so that Eq. (6) yields $l_c > 0$; that is, the reconstruction tendency does change at a finite critical thickness. Since we are interested in the cases where reconstruction tendency changes ($l_c > 0$), we also examine the *opposite* combination of $\frac{n}{m} > 1$ and $\Delta E_S > 0$. 4d noble-metal (001) surfaces fall into this category. It is known that while the bulk Ir (001) surface exhibits the (5×1) reconstruction, which increases the surface atomic density by 20%, 4d noble metal bulk (001) surfaces do not reconstruct [28]. Therefore, for 4d metals, $\Delta E_S > 0$. The more complex Pt(001) and Au(001) reconstructions [17,18] are not considered in the present work, as they require supercells too large for routine DFT simulations.

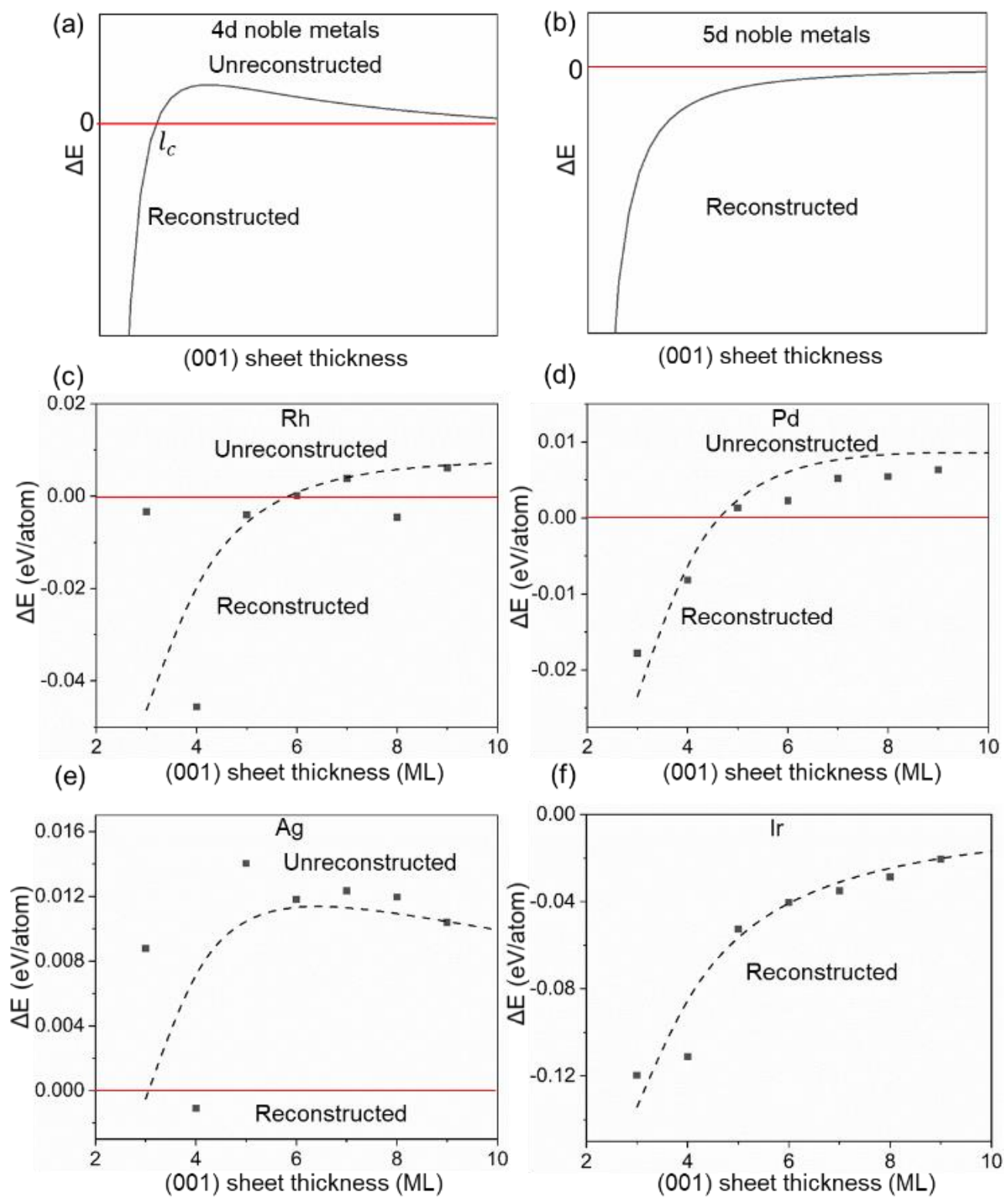

FIG. 3. (a), (b) Schematic average energy difference between (5×1)-reconstructed and unreconstructed structures as a function of (001) sheet thickness for 4d and 5d noble metals, respectively. (c-f) $\Delta E$ versus sheet thickness for Rh, Pd, Ag and Ir, respectively. The black dots are values obtained directly from total-energy calculations, while the dashed curves are derived from surface energies.

As Fig. 3 shows, the $\Delta E$-thickness curves for the 4d noble-metal (001) sheets (Rh, Pd, and Ag) indeed cross zero. That is, for sufficiently thin 4d metal sheets, the unreconstructed structure is no longer the ground state; instead, the (5×1)-reconstructed structure is energetically more stable. The more complex reconstructions, namely, Au(001)-*c*(28×48) and Pt(001)-*c*(26.6×118), are not considered in the present work. For these two structures, the relations $\frac{n}{m} > 1$ and $\Delta E_S > 0$ also hold for the 4d noble metals (thus $l_c > 0$). Therefore, while we conclude that surface reconstruction must be induced, the question of which reconstructed structure is stable for thin 4d noble-metal (001) sheets remains open, since other superstructures including the *c*(28×48) and *c*(26.6×118) reconstructions may compete with the (5×1) structure. Compared with the (110)-(1×2) case (Fig. 2), the deviation between the directly calculated $\Delta E$ and the derived $\Delta E$ in Fig. 3 is generally larger. This primarily arises from the stronger quantum oscillations of the thinner sheets shown in Fig. 3. Nevertheless, the critical thickness $l_c$ derived from surface energies and the actual critical thickness $l_c^*$ agree very well (Table I), indicating that quantum oscillations (and also lattice strains) have only a minor effect on the reconstruction tendency of (001) sheets. It was reported that ferromagnetism was induced in unreconstructed Pd (001) sheets with particular thicknesses of 1, 4, 9, and 15 ML [29]. In our study, Pd(001) sheets with thicknesses of 3 and 4 ML change the reconstruction tendency (from unreconstructed to reconstructed, Fig. 3(d)). Therefore, we study the magnetic properties of the four-layer sheet to assess the effects of magnetism on changing reconstruction tendency. When the in-plane lattice constants are fixed to those of the bulk Pd (001) surface, as was done in ref. [29], our PBEsol calculations also obtain the ferromagnetic ground state for the unreconstructed structure. The calculated magnetic moment is 0.162 $\mu_B$ per surface Pd atom, comparable with the reported value of ~0.14 $\mu_B$ at the LSDA level [29]. However, when the structure is fully optimized (as throughout this work), resulting in a 3.3% contraction of the in-plane lattice constant, the magnetism vanishes—all atomic magnetic moments become negligible. We also investigate the possible magnetism of the fully optimized reconstructed four-layer sheet, which likewise turns out to be nonmagnetic. Therefore, we conclude that the effects of magnetism on reconstruction tendency of Pd(001) sheets are negligible. For 5d noble metals, $l_c < 0$, so no change in the reconstruction tendency is expected (Fig. 3(b)). Ir(001) exemplifies this case (Fig. 3(f)): the sheets remain reconstructed, irrespective of sheet thickness.

According to Eq. (6), $l_c > 0$ (reconstruction tendency changes) requires $\left(\frac{n}{m} - 1\right)/E_S > 0$. That is, either $n > m$ and $\Delta E_S > 0$ (for example, the case of 4d noble metal (001) quasi-hexagonal reconstruction) or $n < m$ and $\Delta E_S < 0$ (for example, the case of 5d noble metal (110) missing-row reconstruction). In the former case, the bulk surfaces do not reconstruct ($\Delta E_S > 0$), whereas sufficiently thin nanosheets transform into reconstructed structures that increase the surface atomic density ($n > m$). In the latter case, the bulk surfaces do reconstruct ($\Delta E_S < 0$), whereas sufficiently thin sheets transform into unreconstructed structures that also increase the surface atomic density ($n < m$). Therefore, in both cases, the change in reconstruction tendency increases surface atomic density of 2D metals (Fig. 4).

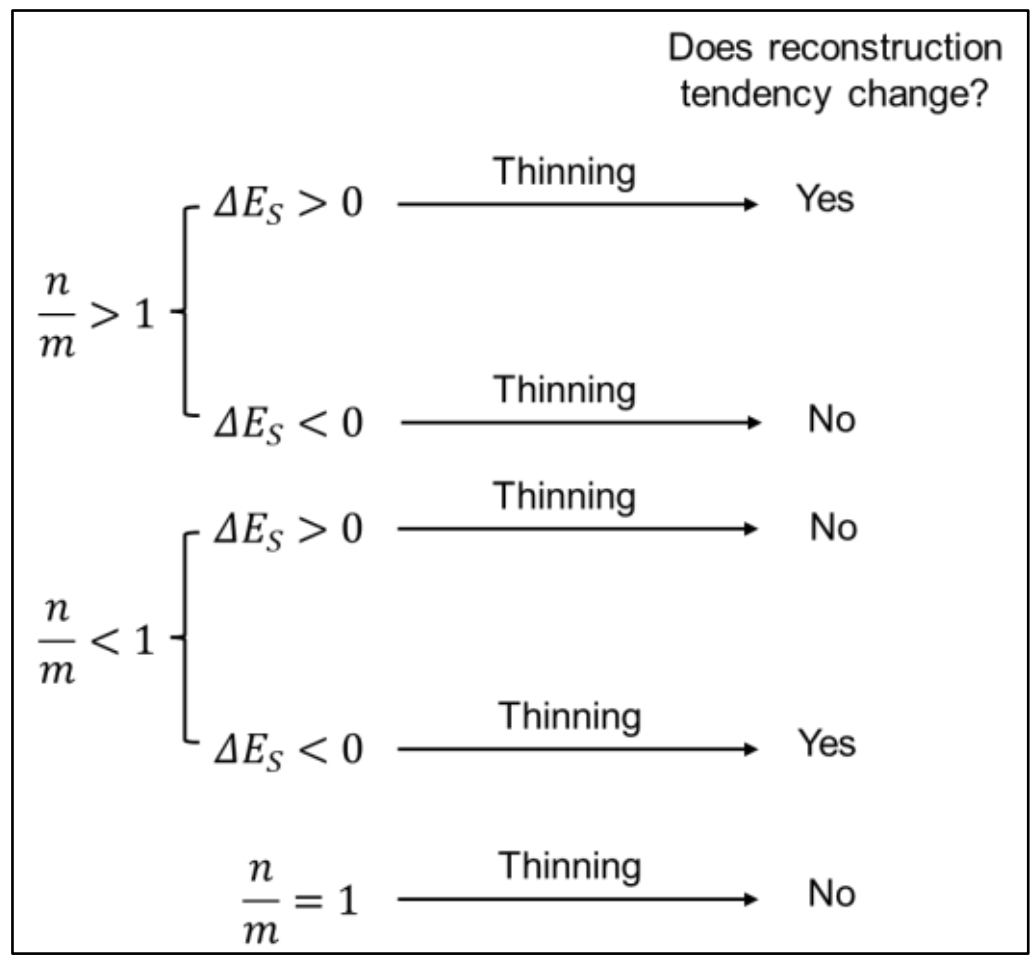

FIG. 4. Summary of the change in reconstruction tendency of 2D metals upon thinning,

showing that thinning always favors structural changes that increase surface atomic density. $n$ and $m$ are the numbers of atoms in a bulk atomic layer and in the reconstructed surface layer within the same lateral area, respectively. $\Delta E_S$ is the surface energy difference between the reconstructed and unreconstructed bulk surfaces; $\Delta E_S < 0$ if the reconstructed surface is more stable, and $\Delta E_S > 0$ otherwise.

Finally, we comment on the case in which the surface atomic density remains unchanged ($n = m$) upon reconstruction, although this situation is rare for noble metals, if it occurs at all. In contrast, non-noble metal surfaces such as W(001) [30] and Mo(001) [31] do exhibit such reconstructions. In this case, $l_c = 0$ (Eq. (6)). Consequently, 2D metal sheets are expected to retain the reconstruction tendencies of the corresponding bulk surfaces regardless of thickness, unless surface stress or quantum oscillations cause the actual $\Delta E$ to deviate so substantially from the derived value that it crosses zero. An overview of the change in reconstruction tendency of 2D metals upon thinning is given in Fig. 4.

In summary, we have established a unified thickness criterion for surface reconstruction in ultrathin metal sheets. For sufficiently thin nanosheets, the (1×2) reconstruction of 5d noble-metal (110) sheets is lifted, whereas the quasi-hexagonal reconstruction of 4d noble-metal (001) sheets is induced. In both cases, thinning drives the structural change that increases the surface atomic density. This result unifies two apparently opposite thickness dependences within a single thermodynamic picture. The critical thickness at which the reconstruction tendency reverses is captured quantitatively by Eq. (6), which takes only the surface energies and atomic densities of the corresponding bulk surfaces as input, with direct DFT calculations confirming that surface stress and quantum oscillations introduce only minor corrections. The reconstruction behavior of an ultrathin metal sheet can thus be predicted a priori, without any calculation of the sheet itself. Since this density criterion is not specific to noble metals, it provides a unifying link between the well-established reconstruction behavior of bulk metal surfaces and the structural stability of two-dimensional metals, and it offers a baseline for understanding adsorbate-induced reconstructions on the metal nanosheets now emerging in catalysis, bioimaging, and sensing.

We express great appreciation for the financial support from the National Key Research and Development Program of China (2022YFA1504704).

[1] S. Titmuss, A. Wander, and D. A. King, Chemical Reviews **96**, 1291 (1996).
[2] C.-H. Cui and S.-H. Yu, in *Nanotechnology for Sustainable Energy* (American Chemical Society, 2013), pp. 265.
[3] C.-H. Yeh, Y.-J. Lin, Y. I. A. Reyes, Y.-H. Huang, C. Coluccini, and H.-Y. T. Chen, Applied Surface Science Advances **34**, 101016 (2026).
[4] M. J. Yacaman and P. Schabesretchkiman, Surface Science **144**, L439 (1984).
[5] K. Takayanagi, Progress of Theoretical Physics Supplement **106**, 249 (1991).
[6] J. L. F. Da Silva, H. G. Kim, M. J. Piotrowski, M. J. Prieto, and G. Tremiliosi-Filho, Physical Review B **82**, 205424 (2010).
[7] Z. Wang, W. An, Y. Sun, and M. S. Hybertsen, The Journal of Physical Chemistry C **123**, 29783 (2019).
[8] C. J. Owen, Y. Xie, A. Johansson, L. Sun, and B. Kozinsky, Nature Communications **15**, 3790 (2024).
[9] Y. Ma, B. Li, and S. Yang, Materials Chemistry Frontiers **2**, 456 (2018).
[10] S. Yu, C. Zhang, and H. Yang, Chemical Reviews **123**, 3443 (2023).
[11] F. Ren *et al.*, Advanced Materials **n/a**, e12683 (2025).
[12] S. Kashiwaya, Y. Shi, J. Rosen, and L. Hultman, 2D Materials **12**, 033001 (2025).
[13] X. Huang *et al.*, Nature Communications **2**, 292 (2011).
[14] Y. Zhao, X. Tan, W. Yang, C. Jia, X. Chen, W. Ren, S. C. Smith, and C. Zhao, Angewandte Chemie International Edition **59**, 21493 (2020).
[15] M. İ. Törehan Balta and Ç. Kılıç, Modelling and Simulation in Materials Science and Engineering **22**, 025009 (2014).
[16] S. Polsterová, M. Friák, M. Všianská, and M. Šob, Nanomaterials **10**, 767 (2020).
[17] R. Hammer, A. Sander, S. Förster, M. Kiel, K. Meinel, and W. Widdra, Physical Review B **90**, 035446 (2014).
[18] R. Hammer, K. Meinel, O. Krahn, and W. Widdra, Physical Review B **94**, 195406 (2016).
[19] A. Schmidt, W. Meier, L. Hammer, and K. Heinz, Journal of Physics: Condensed Matter **14**, 12353 (2002).
[20] J. J. Schulz, M. Sturmat, and R. Koch, Physical Review B **62**, 15402 (2000).
[21] P. Hohenberg and W. Kohn, Physical Review **136**, B864 (1964).
[22] W. Kohn and L. J. Sham, Physical Review **140**, A1133 (1965).
[23] G. Kresse and J. Hafner, Physical Review B **47**, 558 (1993).
[24] J. P. Perdew, A. Ruzsinszky, G. I. Csonka, O. A. Vydrov, G. E. Scuseria, L. A. Constantin, X. Zhou, and K. Burke, Physical Review Letters **100**, 136406 (2008).
[25] L. Vega, J. Ruvireta, F. Viñes, and F. Illas, Journal of Chemical Theory and Computation **14**, 395 (2018).
[26] S. Olivier, G. Tréglia, A. Saúl, and F. Willaime, Surface Science **600**, 5131 (2006).
[27] M. I. Haftel and K. Gall, Physical Review B **74**, 035420 (2006).
[28] V. Fiorentini, M. Methfessel, and M. Scheffler, Physical Review Letters **71**, 1051 (1993).
[29] S. C. Hong, J. I. Lee, and R. Wu, Physical Review B **75**, 172402 (2007).
[30] C. L. Fu, A. J. Freeman, E. Wimmer, and M. Weinert, Physical Review Letters **54**, 2261 (1985).
[31] R. S. Daley, T. E. Felter, M. L. Hildner, and P. J. Estrup, Physical Review Letters **70**, 1295 (1993).